\documentclass[pre,aps,twocolumn,superscriptaddress]{revtex4}
\usepackage{amsmath,bm}

\usepackage{graphicx}

\usepackage[latin1]{inputenc}
\newcommand{\beq}{\begin{equation}}
\newcommand{\eeq}{\end{equation}}
\newcommand{\bea}{\end{eqnarray}}

\def\Re{${R}\mkern-1mu e$} 
\def\RE{{R}\mkern-1mu e} 
\def\DE{{ D}\mkern-1mu e} 

\def\sb#1{_{\rm{#1}}}
\def\Sb#1{\sb{\scriptscriptstyle{#1}}}
\def\sp#1{^{\rm{#1}}}
\def\Sp#1{\sp{\scriptscriptstyle{#1}}}
\def\<{\left\langle}
\def\>{\right\rangle}
\def\1{\bm\delta}
\let\^\widehat
\let\-\overline

\newcommand{\Ref}[1]{(\ref{#1})}
\newcommand{\REF}[1]{Eq.~(\ref{#1})}
\newcommand{\BE}{\begin{equation}}    
\newcommand{\EE}{\end{equation}}
\newcommand{\BSE}{\begin{subequations}}    
\newcommand{\ESE}{\end{subequations}}
\newcommand{\BEA}{\begin{eqnarray}}    
\newcommand{\EEA}{\end{eqnarray}} 
\newcommand{\EEa}{\end{eqnarray*}}   
\def\nn {\nonumber} \def\br {\\ \nonumber} 
\newcommand{\B}[1]{{\bm{#1}}}
\newcommand{\C}[1]{{\mathcal{#1}}}    
\newcommand{\BC}[1]{\bm{\mathcal{#1}}}
 \def\r{\B {r}} 
 \newcommand{\ve}{\varepsilon} 
\newcommand{\p}{\partial}           

\def\<{\left \langle} \def\>{\right\rangle}

\begin{document}
\title{Additive Equivalence in  Turbulent Drag Reduction by Flexible and 
Rodlike Polymers}
\author{Roberto Benzi}
\affiliation{Dip. di Fisica and INFN, Universit\`a ``Tor
Vergata", Via della Ricerca Scientifica 1, I-00133 Roma, Italy}
\author{Emily S.C. Ching}
\affiliation{Dept. of Physics, The Chinese University of Hong
Kong, Shatin, Hong Kong}
\author{T. S.  Lo}
\affiliation{Dept. of Chemical Physics, The Weizmann Institute
of Science, Rehovot 76100, Israel}
\author{Victor S.  L'vov}
\affiliation{Dept. of Chemical Physics, The Weizmann Institute
of Science, Rehovot 76100, Israel}
\author{Itamar Procaccia}
\affiliation{Dept. of Chemical Physics, The Weizmann Institute
of Science, Rehovot 76100, Israel}
\begin{abstract}
We address the ``Additive Equivalence" discovered by Virk and
coworkers: drag reduction affected by flexible and rigid rodlike polymers
added to turbulent wall-bounded flows is limited from above by a
very similar Maximum Drag Reduction (MDR) asymptote. Considering
the equations of motion of rodlike polymers in wall-bounded
turbulent ensembles, we show that although the microscopic
mechanism of attaining the MDR is very different, the macroscopic
theory is isomorphic, rationalizing the interesting experimental
observations.
\end{abstract}
\maketitle
\section{Introduction}

Turbulent flows in a channel are conveniently discussed for fixed
pressure gradients $p'\equiv -\partial p/\partial x$ where $x$,
$y$ and $z$ are the lengthwise, wall-normal and spanwise
directions respectively \cite{79MY}. The length and width of the
channel are usually taken much larger than the mid-channel height
$L$, making the latter a natural re-scaling length for the
introduction of dimensionless (similarity) variables. Thus the
Reynolds number \Re, the normalized distance from the wall $y^+$
and the normalized mean velocity $V^+(y^+)$ (which is in the $x$
direction with a dependence on $y$ only) are defined by
\begin{equation}
\RE \equiv {L\sqrt{\mathstrut p' L}}/{\nu_0}\ , \  y^+
\equiv {y \RE }/{L} \ , \  V^+ \equiv
{V}/{\sqrt{\mathstrut p'L}} \ , \label{red}
\end{equation}
where $\nu_0$ is the kinematic viscosity.  Drag reduction by
polymers in a channel geometry is bounded by two asymptotes
\cite{75Vir}. One is the von-K\'arm\'an log-law of the wall for
Newtonian fluids
\begin{equation}
\label{LLW} V^+(y^+) =\kappa_{_{\rm K}}^{-1}\ln y^+ + B\,,
\quad{\rm for}~ y^+ \gtrsim 30  \ .
\end{equation}
While the log-law can be derived using several approaches, the
von-K\'arm\'an constant $\kappa\Sb K\approx 0.436 $ and intercept
$B\approx 6.13$ are only known from experiments and simulations
\cite{79MY,97ZS}. The second asymptote is the Maximum Drag
Reduction (MDR) where the velocity field assumes another log-law
of the form
\begin{equation}
V^+(y^+) = \frac{1}{\kappa_{_{\rm V}}}\ln\left(e\, \kappa_{_{\rm
V}} y^+\right)\, \quad{\rm for}~ y^+ \gtrsim 10
 \ . \label{final}
\end{equation}
This law, which had been discovered experimentally by Virk (and
hence the notation $\kappa_{_{\rm V}}$) was derived theoretically
for {\em flexible} polymers in
\cite{04LPPT,04DCLPPT,04LPPTa,04BDLPT}. The actual velocity
profile in the presence of polymers is bounded between these two
asymptotes:  for sufficiently high values of \Re~ and
concentration of the polymer, the velocity profile in a channel is
expected to follow the law (\ref{final}). For finite \Re~, finite
concentration and finite extension of the polymers one expects
cross-overs back to a velocity profile parallel to the law
(\ref{LLW}), but with a larger mean velocity (i.e. with a large
value of the intercept $B$). The position of the cross overs are
not universal, and they are understood fairly well
\cite{04BDLPT,04BLPT}.
 
In this paper we address the experimental finding that {\em rigid rodlike}
polymers appear to exhibit a very similar MDR (\ref{final}) as
flexible polymers \cite{97VSW}. Since the bare equations of motion
of rodlike polymers differ quite significantly from those of
flexible polymers, one needs to examine the issue carefully to
understand this similarity, which was termed by Virk ``Additive
Equivalence". The aim of this paper is to understand this Additive
Equivalence on the basis of the equations of motion.

In Sec. \ref{s:drag} we address the theory of drag reduction by rigid 
polymers.  In  Sec. \ref{ss:Basic}  we consider the 
equations of motion of rigid polymers (or fibers) in the presence of 
strong wall-bounded turbulence. We explain the interesting differences 
between the interaction of flexible and rodlike polymers with turbulent 
fluctuations. In Sec. \ref{ss:equations} we discuss the difference in 
statistics between flexible and rodlike polymers near thermodynamic 
equilibrium. Section~\ref{ss:turbulence}
is devoted to the statistics of rodlike polymers in turbulent flows with 
a strong shear. In  Sec.~\ref{ss:rules} we address the important issue of how to evaluate the various 
cross-correlation functions between the polymer conformation tensor and the turbulent 
fluctuations. These objects have a seminal role in the theory of drag reduction by rodlike 
polymers. In the following Secs.~\ref{ss:moment} and \ref{ss:energy}
we demonstrate that in spite of the very significant microscopic 
differences between flexible and rodlike polymers, the balance equations 
for momentum and energy have exactly the same form as the corresponding 
equations for flexible polymers. The ``additive equivalence" follows 
from this observation. In Sect. \ref{conclusions} we summarize the paper 
and discuss further the correspondence between drag reduction by 
flexible and rodlike polymers.

\section{\label{s:drag}Theory of Drag Reduction by rodlike polymers}
\subsection{\label{ss:Basic}Basic equation of motion}
\subsubsection{\label{sss:NSE}
Hydrodynamic equations for the polymeric solutions}
The hydrodynamic equations for an incompressible fluid velocity
$\bm{U}\equiv\bm{U}(t,\bm{r})$ in the presence of rodlike polymers
have the form
\begin{subequations}\label{NS}
\begin{eqnarray}\label{NS-a}
  \frac{D\bm U}{Dt} &=& \nu_0\Delta\bm U -\bm\nabla p
    +\bm\nabla\cdot\bm\sigma
\,,\\\label{NS-b}
 0&=&  \bm\nabla\cdot\bm U \ .
\end{eqnarray}
\end{subequations}
Here the fluid density is set to unity ($\varrho\equiv1$), $D/Dt$ is
the substantial derivative
\begin{equation}\label{DDt}
  \frac{D}{Dt} \equiv \frac{\partial}{\partial t}
  +\bm\nabla\cdot\bm U
\,,\end{equation} 
$p\equiv p(t,\bm{r} )$ is the pressure field,  $\nu_0$ is the
kinematic viscosity of the carrier fluid and  ${ \bm \sigma}
\Rightarrow \sigma_{ab}$ is an extra stress tensor caused by the
polymers.

The calculation of the  tensor $\bm \sigma$ for rigid rods is
offered in the literature \cite{88DE}, subject to the assumptions
that the rodlike polymers are mass-less and having no inertia. In
other words, the rodlike polymers are assumed to be at all times in
local rotational equilibrium with the velocity field. Thus the
stress tensor does not have a contribution from the rotational
fluctuations against the fluid, but rather only from the velocity
variations along the rodlike object. Such variations lead to ``skin
friction", and this is the only extra dissipative effect that is
taken into account. The result of these considerations is the
following expression for the additional stress tensor:
\begin{equation}
\sigma_{ab} = 6\nu\sb p\,  n_a n_b \left( n_i n_j \C S_{ij}\right)
\,,\quad \mbox{rodlike polymers}\,,  \label{def-sigma}
\end{equation}
where $\nu\sb p$ is the polymeric contribution to the viscosity at
vanishingly small and time-independent shear;  $\nu\sb{p}$
increases linearly with the polymer concentration, making it an appropriate
measure for the polymer's concentration.  The other
quantities in Eq.~(\ref{def-sigma}) are the velocity gradient
tensor 
\BE \label{def-S} \BC S\equiv(\bm\nabla\bm{U})\Sp{T} \ \Rightarrow
\ \C S_{ab}= \p U_a/\p x_b\,,
\EE
 and $\bm{n}\equiv\bm{n}(t,\bm{r})$  is a unit
($\bm{n}\cdot\bm{n}\equiv1$) director field that describe the polymer's
orientation. Notice, that for flexible polymers the equation for $\sigma_{ab}$ is
completely  different from \REF{def-sigma}:
\begin{equation}
\sigma_{ab} = \nu\sb p\gamma\sb p\,  n_a n_b \,,\quad
\mbox{flexible polymers}\ . \label{def-sigma-1}
\end{equation}
Here $\gamma\sb p$ is the polymeric relaxation frequency. The
difference between Eqs.~\Ref{def-sigma} and \Ref{def-sigma-1} for
$\sigma_{ab}$ for the rodlike and flexible polymers reflects their
very different microscopic dynamics. For the flexible polymers the
main source of interaction with the turbulent fluctuations is the
stretching of the polymers by the shear.  This is how energy is
taken from the turbulent field, introducing an additional channel
of dissipation without necessarily increasing the local gradient.
In the rigid case the dissipation is only taken as the skin
friction along the rodlike polymers. Having in mind  all these
differences it becomes even more astonishing that the macroscopic equations
for the mechanical momentum and kinetic energy balances  are
isomorphic for  the rodlike  and flexible polymers, as is demonstrated
below.

\subsubsection{\label{sss:rot-eq}Equation for the orientation
field of dilute rods solution} 
The equation for $\bm{n}(\r,t)$ has the form
\begin{equation}
\label{poleq}
  \frac{D\bm n}{Dt} = \left(\1 -\bm n\bm n\right)\cdot\BC S\cdot\bm n\
  \Rightarrow\
  \frac{D n_a}{Dt} = \left(\delta_{ai} - n_a n_i\right) \C S_{ij}
  n_j \ .
\end{equation} 
This equation conserves the unit length of the director $\bm{n}$.
The theory is written more compactly in terms of the (normalized) conformation tensor
\begin{equation}\label{R}
 \BC R\equiv \B n \B n\ \Rightarrow \C R_{ab}\equiv n_a n_b\ .
\end{equation}%
The equation of motion of this object follows from \REF{poleq}:
\BE\label{R-eq}
 \frac{D \C R_{ab}}{Dt}=
\C S_{ai}\C R_{ib}+\C S_{bi}\C R_{ia}-2\,\C R_{ab} (\C S_{ij}\C
R_{ij}) \ .
\end{equation}
In terms of $\BC R$ we can rewrite $\B \sigma$, \REF{def-sigma}, as 
\begin{equation}\label{polstress1-a}
\bm\sigma = 6\nu\sb p\BC R  \,\mbox{Tr}\{\BC R\cdot \B S\}\
\Rightarrow\   \sigma _{ab} = 6\nu\sb p\, \C R_{ab}( \,\C R_{ij}
\C S_{ij}) \ .\end{equation}

\subsection{\label{ss:equations}The statistics of rodlike polymers
near thermodynamic equilibrium} 
In order to understand the interaction of rodlike polymers with wall
bounded turbulent fluctuations, we need to start along the well
charted road of dynamics near equilibrium, and examine the
solution of these equation for strong shears.

\subsubsection{\label{sss:eq-near-eq} Equations of motion near
thermodynamic equilibrium} 
Consider the shear $\BC S(\r,t)$ in \REF{poleq} as
independent of space and time, and replace $\BC S(\r,t) \Rightarrow \B S$.
Near equilibrium rodlike polymer (or fibers) experience rotational
disorder due to local thermal velocity fluctuations that can be
considered as a Brownian motion in the space of angles. As usual, the
effect of thermal fluctuations can
be mimicked by adding to the RHS of \REF{poleq} a Langevin random
force $\B f(t)$.  When the rigid polymer is symmetric the
hydrodynamics can depend only on the dyadic product $\B n\B n$.
 By pre- and post-multiplying the 
equation above by $n_b$, and taking average over the Brownian fluctuations, 
we have the transport equation for the second moment $R_{ab}\equiv \overline{n_a n_b}$,
where and overbar indicates an average over the polymer configurations:
\begin{equation}
\label{eq4} \frac{D \overline{\C R}_{ab}}{Dt}= S_{ai} \overline{\C
R}_{ib}+S_{bi}\overline{\C R}_{ia}-2S_{ij}\overline{\C
R}_{ij}\overline{\C R}_{ab}- 6\gamma\Sb B \left(  \overline{\C R}_{ab}-\frac{\delta_{ab}}{3 } \right)\,,
\end{equation}
where $\gamma\Sb B$ is the Brownian rotational frequency, proportional to the temperature.
The derivation of this equation in the literature employs the closure
\begin{equation}
\label{eq5} \overline{\C R_{ij} \C R_{ab}}\, S_{ab}=\overline{\C
R}_{ij}\overline{\C R}_{ab}\, S_{ab}\,,
\end{equation}
which is rationalized in \cite{88DE}
\subsubsection{\label{sss:SSS}Solution for a simple shear flow}
For a simple shear  the velocity gradient satisfies
\BE
\label{SS}
S_{ab}=S \delta_{ax} \delta_{by}\,,
\EE
\REF{eq4} in the stationary, space-homogeneous case turns
into: \BE\label{R-eq3}
\DE\sb r\,\left(\delta_{ax}\overline{\C R}_{yb}+\delta_{bx}\overline
 {\C R}_{ya}-2\, \overline{\C R}_{ab} \overline{\C R}_{xy}
\right)= 2 (3\, \overline{\C R}_{ab}- \delta_{ab})\ .
\EE
Here we introduces the  Deborah number for the rod-like polymers:
\BE\label{De} 
\DE\sb r\equiv S/ \gamma\Sb B\ .\EE 
Equation (\ref{R-eq3}) was solved in the limit $\DE\sb r\gg 1$ \cite{88DE} with the final results
\BEA 
\nn
\overline{\C R}_{xx}&\approx & 1\gg \overline{\C R}_{xy}\approx
\frac{1}{{2\DE\sb r}^{1/3}}\gg
\overline{\C R}_{yy}\approx \frac{2^{1/3}}{\DE\sb r^{2/3}}\,, \\
 2 \, \overline{\C R}_{xy}^{\;2}&\approx&
 \overline{\C R}_{xx}\overline{\C R}_{yy}\ .
\label{orders}\EEA 

It is interesting to  compare the statistics of rodlike and flexible
polymers in strong laminar shears. For example (cf. \cite{04LPPTa} and references therein),
\begin{eqnarray}
\label{dif-r} \overline{\C R}_{xy}&\sim& \DE\sb r^{1/3}\overline{\C R}_{yy}\,, \quad \mbox{rodlike}\,,\\
\label{dif-f} \overline{\C R}_{xy}&\sim& \DE\sb f \ \overline{\C
R}_{yy}\,, \quad \mbox{flexible}\ . 
\end{eqnarray}
Here the  Deborah number for flexible polymers $\DE\sb f$ is
defined with the flexible polymer relaxation frequency $\gamma\sb p$,  $\DE\sb f = S/ \gamma \sb p$.
The  different  dependence on Deborah number  stems from the
very different microscopic dynamics that leads to different expressions
for the polymeric stress tensor $\sigma_{ab}$ for the rodlike and
flexible polymers.
\subsection{\label{ss:turbulence}
 Statistics of rigid rods in turbulent flow with strong shear} 
 In the presence of turbulence the fluctuations are no longer
thermal and  the statistical description of the polymer orientation
is accomplished with the mean values
\BE \label{turb-mean} 
R_{ab}\equiv \< \C R_{ab}\>  \  ,\EE 
where the angular brackets denote an average over the turbulent fluctuations. 
For well developed turbulence  it is expedient to use the Reynolds decomposition
in which the velocity field $\B U(\B r)$ is written as a sum of
its average (over time) and a fluctuating part:
\begin{equation}
\B U(\B r,t) = \B V(y) + \B u(\B r,t) \ , \quad \B V(y)
\equiv \langle \B U(\B r,t) \rangle \ .
\end{equation}
For a channel of large length and width all the averages, and in particular
$\B V(y) \Rightarrow V(y)$, are functions of $y$ only.
Correspondingly the shear is written as a sum of the mean shear  $S_{ab}$ and fluctuating
 shear $\B s(\r,t)$ with zero mean:
\BE\label{decom1}
\B S=\<\BC S(\r,t)\>\,, \quad  \BC S(\r,t)= \B S + \B s(\r,t)\,,
\quad \< s(\r,t)\>=0\,,
\EE
With these notations and  after averaging over the statistics of
turbulence,  \REF{R-eq} takes the form 
 \BE\label{turb-a}
\left \langle  \frac{D \C R_{ab}}{Dt}\right \rangle = A_{ab}-B_{ab} \ ,
\EE
where $A_{ab}$ contains all the terms in which the mean shear appears explicitly,
and $B_{ab}$ contains only the fluctuating part of the shear:
\begin{eqnarray}
A_{ab} &\equiv& S_{ai}R_{ib}+S_{bi}R_{ia}-2\,S_{ij} \<\C  R_{ab} \C R_{ij}\>  \ , \nonumber\\
B_{ab}&\equiv&2 \<\C R_{ab} (s_{ij}\C R_{ij})\>  -\<s_{ai}\C
R_{ib}\>-\<s_{bi}\C R_{ia}\> \  . \label{AB}
\end{eqnarray}
In a steady state in channel geometry the LHS of Eq. (\ref{turb-a}) contains only one non-vanishing
terms, which is $\langle  \B u\cdot \B\nabla \BC R\rangle$. This term is responsible for the turbulent part 
of the physical flux of $\BC R$. In all our derivation below we will assume that such terms are
small compared to all 'local' contributions to the balance equations for the momentum and
energy. This assumption will have to be tested a-posteriori.   

With this in mind Eq. (\ref{turb-a}) reads simply $A_{ab}=B_{ab}$. This will allow us
to estimate the crucial correlations functions that appear in the theory of drag reduction below. The 
LHS of this relation can be made explicit in channel geometry; using Eq. (\ref{SS}) we find

\BE\label{R-turb-a}
A_{ab}= S\,\left(\delta_{ax}R_{yb}+\delta_{bx} R _{ya}-2\, \< \C R_{ab}  \C R 
_{xy}\>
\right) \ .
\EE
Using definition \Ref{R} and constraint $|\B n|=1$ this Eq. can be
rewritten in components as: 
\BSE \label{SSS} 
\begin{eqnarray} 
\label{SSS:xx}
A_{xx}&=& 2\, S \< \C R_{xy} (\C R _{yy}+\C R_{zz})\> \,, \\ 
\label{SSS:yy}
A_{yy}&=&  -2\,S \,  \< \C R _{xy}  \C R _{yy}\>  \,,\\
\label{SSS:zz}
A_{zz}&=&   -2\,S \,  \< \C R _{xy}  \C R _{zz}\>  \,,\\ 
 \label{SSS:xy} A_{xy}&=&  S \,\< \C R_{yy}(1-2\, \C R _{xx})\> \ .
\end{eqnarray}
\ESE 

In writing down expressions for $B_{ab}$ we will make explicit use of the expected solution
for the conformation tensor in the case of large mean shear, $S^2\gg \langle s^2\rangle$.
In such flows we expect a strong alignment of the rod-like polymers along the streamwise
direction $x$. The director components $n_y$ and $n_z$ are then much smaller than 
$n_x\approx 1$. For large shear we can expand $n_x$ according to 
\BE\label{exp}
n_x=\sqrt{1-n_y^2-n_z^2}\approx 1-\frac12(n_y^2+n_z^2)\,,  
\EE
We note that for $n_x=1$ (when the shear is actually infinite) the object $B_{ab}$ vanishes
since $\langle \B s\rangle =0$. We therefore represent $B_{ab}$ in order on $n_y\sim n_z$,
keeping up to second order. We will show below that it is important to keep terms of second order
since some of them have the same magnitude as terms which are formally of first order in the
smallness. All terms of third order are smaller in magnitude than the terms that we keep.   The
first two orders read:
\BSE\label{div1}
\BEA
B^{(1)}_{xx} &=&2\,\<s_{xy}\C R_{xy}+s_{xz}\C R_{xz} \> 
\,,\label{div-xx1} \\
B^{(1)}_{yy}&=& -2\,\<s_{xy}\C R_{xy}\>\,, \label{div-yy1}\\
B^{(1)}_{zz}&=& -2\,\<s_{xz}\C R_{xz}\>\,, \label{div-zz1}\\
B^{(1)}_{xy}&=& \< \C R_{xy}(s_{xx}-s_{yy})-\C R_{xz} s_{yz} \>\,, 
\label{div-xy1}
   \EEA
   \ESE \BSE\label{div2} \BEA  \nn B^{(2)}_{xx} &=&2\,\<\C
   R_{yy}(2s_{yy}-s_{zz}) +\C R_{zz}(2s_{zz}-s_{xx})\right.\\&&
 \left. +2 \C R_{yz}(s_{yz}+s_{zy}) \> \,,\label{div-xx2} \\
B^{(2)}_{yy}&=&\<\C R_{yy}(s_{xx}-2s_{xy})- 2\, \C R_{xz} s_{xz})\>\,,
   \label{div-yy2} \\ \label{div-xy2}
B^{(2)}_{xy}&=& \< \C R_{yy}(s_{xy}+3 s_{yx})+ \C R_{zz}s_{yz}\right. 
\\ \nn
  &&\left. +\C R_{yz}(s_{xz}+s_{zx})\>\ .
   \EEA \ESE 
   Equations \Ref{SSS}, \Ref{div1} and \Ref{div2}
serve as basis for further analysis.
\subsection{\label{ss:rules}Closures and orders of magnitude}
\subsubsection{\label{sss:stat-ob}Statistical objects of interest}

A theory of turbulent channel flows of Newtonian fluids can be constructed in terms of the mean shear $S(y)$, the Reynolds stress $W(y)$ and
the kinetic energy $K(y)$; these are defined respectively as
\begin{equation}\label{def-KW}
S(y)\equiv d V(y)/d y \ , \  W (y)\equiv - \langle u_xu_y\rangle
\ , \  K(y) =  \langle |\B u|^2\rangle/2  .
\end{equation}
In the rodlike polymer case the additional stress tensor
$\sigma_{ij}$ and its various correlation functions needs to be
considered as well. For that purpose we turn now to the analysis
of the necessary statistical objects.

First note that the 
expansion~\Ref{exp} allows us to express  all products $\C
R_{ab}\C R_{cd}=n_a n_b n_c n_d$ in terms that are linear in $\BC R$, up to third
order terms in $n_y\sim n_z$. For
example, \BEA\label{exm} \C R_{xx}^2&\approx&  1- 2(\C R_{yy}+\C
R_{zz})\,, \quad \C R_{xy}^2 \approx \C R_{yy}\,,\br \C R_{xy}\C
R_{xz}&\approx& \C R_{yz}\,, \quad \C R_{yy}\C R_{ij}\approx \C
R_{yy}\delta_{ix}\delta_{jx}\,, \qquad \mbox{etc.} \EEA Actually
we have used these estimates in the derivation of Eqs.~\Ref{div1}
and \Ref{div2}.

As a further preparation for the theory below we analyze various
statistical objects in the turbulent environment and estimate
their magnitudes. Based on experimental observations and DNS data
we assume that statistics of turbulent fluctuations  does not deviate too much
from isotropy. Explicitly,
\BE\label{expect}
\<s_{xx}^2\>\sim \<s_{yy}^2\> \sim\<s_{zz}^2\>\sim \<s_{xy}^2\>\sim \dots
\EE
Here and below the notation $\sim$ means ``the same order of
magnitude  (i.e. correct to leading order up to coefficients of the
order of unity)".

Second, consider correlation functions of turbulent fluctuation of
the shear, $s_{ij}$, i.e. $\langle s_{ij}s_{k\ell}\rangle$.  At
distance $y$ from the wall the correlation functions is dominated
by eddies of size $y$. Thus
\begin{equation}
\langle s_{ij} s_{k\ell}\rangle\sim K(y)/y^2 \ , \quad \text{for
all}~ ij,\, kl \ ,\label{2nd}
\end{equation}

Third, consider  a cross-correlation functions    of the tensors
$\BC R$ with the turbulent shear $\B s$. For this goal take the
leading terms in the LHS and the RHS of $xy$ \REF{SSS:xy} and
\Ref{div-xy1}:
\BE\label{xy-eq}
|S R_{yy}|\simeq |B^{(1)}_{xy}|\lesssim R_{xy}\frac{\sqrt{K}}y\ ,
\EE 
where on the RHS we have used the Cauchy-Schwartz inequality and \REF{2nd}. The notation $\lesssim$  means ``$\sim$ in the sense defined above, or having smaller order of magnitude". 
 Below, (cf. Sec.~\ref{ss:energy}) we show that this estimate is saturated. Therefore expecting
that $R_{yy}\sim R_{xy}^2$ we have
\BE\label{est1}
S R_{xy}\sim \frac{\sqrt{K}}y\ . 
\EE 
This estimate can be recast into an intuitive form which is
\begin{equation}
R_{xy}= \langle n_x n_y\rangle \sim \frac{\langle\sqrt{ s^2}\rangle}{S} \ .
\end{equation}
This is in direct accord with the understanding that the degree of deviation from perfect
alignment of the rodlike polymers ($R_{xx}=1$) is proportional to the turbulent fluctuations
relative to the mean shear.

Taking now the leading terms in $yy$ \REF{SSS:yy} and \Ref{div-yy1} we 
have
\BE\label{yy-eq}
S \< \C R_{yy}\C R_{xy}\> \simeq -\frac12B^{(1)}_{yy}=
\<s_{xy}\C R_{xy}\>\ . \EE Estimating $ \< \C R_{yy}\C R_{xy}\>$
as $\sim R_{yy}R_{xy}$, and using \REF{est1}, we have
\BE\label{est2a}
\<s_{xy}\C R_{xy}\>\sim R_{yy}\frac{\sqrt{K}}y\ .
\EE
Notice  then  the Cauchy-Schwartz inequality for the same correlation,
\BE\label{est2aa}
\<s_{xy}\C R_{xy}\>\lesssim R_{xy}\frac{\sqrt{K}}y\,,
\EE
 gives much higher upper bound, than the real estimate
\REF{est2a}. This shows how important is the use of the equations of motion in
estimating various correlation functions; sometime the direct Cauchy-Schwartz estimate
saturates, and sometime it is a gross overestimate. The reason for why different
correlations have different order of magnitude can be traced back to Eqs. (\ref{SSS}) which
indicate that the diagonal components of $\B A$ are cubic in the small parameter $n_y\sim n_z$ while
$A_{xy}$ is quadratic. 

Notice that the correlator $\<\C R_{ij}s_{ij}\>$ has the
contributions of the type  of $\<s_{xy}\C R_{xy}\>$ and  $\<s_{yy}
\C R_{yy}\>$. Both of them have the same estimate $R_{yy}\sqrt{K}/
y$.  Therefore we can write \BE\label{est2c} \<\C R_{ij}s_{ij}\>
\sim R_{yy}\frac{\sqrt{K}}y\ . \EE

\subsection{\label{ss:moment}The momentum-balance equation}
At this point we can apply our estimates in the context of the
balance equations for the mechanical momentum and the energy. We
begin with the former, which is exact. It reads
\begin{equation}
\label{balP}
\nu_0 S + W + \langle \sigma_{xy} \rangle = p'(L-y) \ ,
\end{equation}
where $p'\equiv -\partial p/\partial x$. Near the wall $y\ll L$
and the RHS of this equation is approximated as $p'L$, a constant
production of momentum due to the pressure gradient. On the LHS we
have the Reynolds stress which is the ``turbulent" momentum flux
to the wall, in addition to the viscous and the polymeric
contributions to the momentum flux.

Using Eq.~(\ref{def-sigma}) and Reynolds decomposition~\Ref{decom1} we 
compute
\begin{equation}\label{sigma}
\langle \sigma_{xy} \rangle = 6\nu\sb p \langle \C R_{xy}\C
R_{ij}\C S_{ij}\rangle = 6\, \nu\sb p[ S\< \C R_{xy}^2\>+ \< \C
R_{xy}\C R_{ij} s_{ij}\>] \ .\end{equation} With \REF{exm} the
first term in the RHS of this equation can be estimated as
follows: \BE\label{est3} 6\, \nu\sb p S\< \C R_{xy}^2\>= c_1
\nu\sb p R_{yy}S\,, \qquad c_1\simeq 6\ . \EE 
On the other hand, using the estimate (\ref{est2c}) one sees that 
the second term in the RHS of \REF{sigma} is negligible.

Finally we can present the momentum balance equation in the form
\begin{equation}
\label{monabl} \nu_0 S + c_1\nu\sb p  R_{yy}S + W  = p'L  \ .
\end{equation}
Another way of writing this result is in the form of an effective viscosity,
\begin{equation}
\label{eff}
\nu(y)  S + W  = p'L  \ ,
\end{equation}
where the effect of the rodlike polymers is included by the
effective viscosity $\nu(y)$:
\begin{equation}
\label{effnu} \nu(y) \equiv \nu_0 + c_1\nu\sb p R_{yy} \ .
\end{equation}
\subsection{\label{ss:energy} Turbulent Energy Balance Equation}

In considering the balance of energy in a channel flow it pays to separate the spatial
directions, since we can learn separate bits of information from each such equation.
Introduce the partial kinetic energy density
\begin{equation}\label{KEj}
K_a(y)\equiv \frac12 \left\langle u_a^2 \right\rangle\,,\quad
K(y)=K_x+K_y+K_z\ ,
\end{equation}%
 and consider the partial energy balance of  $K_a(y)$ 
\BE\label{bal-def}
\frac{\p K_a(y)}{\p \,t}+ R_a+\ve_a\sp{dis}+\ve_a\sp p=W
(y)S(y)\delta_{ax}\,.
\EE%
The total density of the kinetic energy at given distance $y$ from
the wall is: 
\BEA
 \label{bal-tot} && \frac{\p K(y)}{\p
\,t}+\ve\sp{dis}+\ve\sp p=W (y)S(y)\,, \br &&
\ve\sp{dis}=\sum_a \ve_a\sp{dis}\,, \quad  \ve\sp p=\sum_a
\ve_a\sp p\,, \quad \sum_a R_a  =0\ .
\EEA%
The various symbols in the last two equations are explained as follows:
The RHS of these equations describes the energy flux from the
mean flow to turbulent fluctuations due to the correlation between stream-wise
and cross-stream components of the turbulent velocity, known as
the Reynolds stress $W$, see \REF{def-KW}. Remarkably, in channel geometry 
this flux exists only  in the equation for the streamwise
velocity fluctuations, $K_x$.
 
 The  term $R_a(y)$ is known as the
``Return to isotropy" \cite{Pope}, and it vanishes for isotropic turbulence in which $K_x=K_y=K_z$.
Otherwise it redistributes partial kinetic energy between
different vectorial components and does not contribute to the
total balance~\Ref{bal-tot}.  A simple model for this
term~\cite{Pope} is
\BE\label{isotr} R_a\sim \frac{\sqrt{K}}{y}\Big(K_a-\frac{K}3
\Big)\ . \EE 
As usual  the local ``outer scale of
turbulence" was estimated as the distance to the wall $y$. The 
order of magnitude estimate (\ref{expect}) is in accord with the role of this term 
in returning to local isotropy. 

 The term
\begin{equation}\label{vis-dis}
\ve\sp{dis}_a\simeq \nu_0 \sum _j\left\langle s^2_{ja}
\right\rangle\,, \quad \mbox{no sum over}\ a\ ,
\end{equation}%
on the LHS of \REF{bal-def} is the rate of  the viscous
dissipation,  proportional to the kinematic viscosity of
the carrier fluid $\nu_0$. Lastly, 
the polymer contribution to the energy balance, denoted as $\ve\sp
p_a $, can be exactly computed as
\BE\label{en-bal}
 \ve_a \sp p= \left\langle \sigma_{aj} s_{aj} \right\rangle = 6
\,\nu\sb p \left\langle s _{ai} \C R_{ai}\left(S \, \C R_{xy}+ s
_{jk}\C R_{jk}   \right) \right\rangle\ .
\EE
 Notice that Eqs. \Ref{bal-def} and \Ref{bal-tot} are written
``in the local approximation", in which the energy flux in the
physical space is neglected. This is consistent with neglecting the term 
 $\langle  \B u\cdot \B\nabla \C R\rangle$ in our discussion after Eq. (\ref{AB}). A justification of this approximation in the problem of drag reduction by polymers is found in 
\cite{04LPPT,04DCLPPT}.

Using the expansion~\Ref{exp} we can rewrite the equation for the dissipation
rate $\ve\sp p$ as a series in the small parameter  $n_y\sim n_z$:
\BEA\nn \ve\sp p &=& 6\nu\sb p \< \big\{s_{xx} +\big[\C
R_{xy}(s_{xy}+s_{yx})+ \C R_{xz}(s_{xy}+s_{yx}) \big]_1 \right .\\
\label{exp-1} &&  + \big[s_{yy}\C R_{yy}+s_{yz}\C R_{yz}+\mbox {more}
\big]_2+\dots\big\}\br &&\left . \times \big\{S\C R_{xy}+
s_{xx}+\big[\cdots\big]_1 +\big[\cdots \big]_2+ \dots \big\}\> \ .
\EEA 
In the square brackets $ [\cdots]_1$ we displayed all the terms which are linear in
the small parameter. In the square brackets $ [\cdots]_2$ we show  two
 quadratic  terms, and there are more of them as indicated.  The symbol $+\dots$  stands for ``higher order terms". In the second curly brackets the 
square brackets $ [\cdots]_1$  and $ [\cdots]_2$ are identical to the corresponding
terms in the first curly brackets. 
The two
leading terms in \REF{exp-1} are proportional $\<s_{xx}^2\>$ and $S\< \C R_{xy}
s_{xx} \>$. Using Eqs.~\Ref{2nd},  \Ref{div-xy1} and \Ref{xy-eq} one sees
that both leading terms have the same order of magnitude, $\sim
K(y)/y^2$. In fact we will argue that these two terms must cancel
each other, up to terms of the a higher order contributions $\sim
R_{yy} K(y)/y^2$.

To see that such a cancellation must exist, we note that near the
MDR we expect the polymer contribution to the dissipation to
balance the production term $WS$ (cf. Eq. (\ref{EB}) below). We
will show later that this production term is  $y$ independent
(where $y$ is the distance from the wall). On the other hand we
will show that $K(y)$ is linear in $y$, making $K/y^2$ very large
near the wall. Therefore $WS$ cannot be balanced by $K/y^2$. To
avoid using the final results at this stage, and nevertheless to
see that a cancellation must exist, we can focus just on the
stationary balance \REF{bal-def} for the $y$ component, in which the
RHS is zero. Notice also that near the MDR the polymer
contribution to the energy balance dominates over the viscous
dissipation and the nonlinear energy flux  from large to small
scales\cite{04LPPT,04DCLPPT,04LPPTa}. The latter was evaluated
in~\cite{04LPPT} as $K^{3/2}/y$ which is exactly the evaluation of
the return to isotropy term. The conclusion is that near the MDR
\BE\label{est4}
 \ve_y \sp p= \left\langle \sigma_{aj} s_{aj} \right\rangle = 6
\,\nu\sb p \left\langle s _{yi} \C R_{yi}\left(S \, \C R_{xy}+ s
_{jk}\C R_{jk}   \right) \right\rangle\approx 0\ . \EE
This expression can be  again arranged in orders of magnitude,
similarly to \REF{exp-1}:
\BEA
\nn \ve\sp p_y &=& 6\nu\sb p \< \big\{\big[\C
R_{xy}s_{yx}\big]_{y1} \label{exp-y} + \big[s_{yy}\C
R_{yy}+s_{yz}\C R_{yz} \big]_{y2}+\dots\big\} \right. \br
&&\left . \times \big\{S\C R_{xy}+ s_{xx}+\big[\cdots\big]_1
+\big[\cdots \big]_2+ \dots \big\}\> \ . 
\EEA 
Here the second curly brackets is the same as the second curly brackets in \REF{exp-1}.  
In \REF{exp-y} we find again the same two
large contributions, i.e. $\C R_{xy}s_{yx}\{S\C R_{xy}+ s_{xx}\}+$
l.o.t, where the lower order terms have at least one small factor.
Thus these terms must cancel each other, which means that $S \C
R_{xy}$ cancels $s_{xx}$ inside correlation
functions. But these are exactly the terms that appear in the sum
of the two large terms in Eq. (\ref{exp-1}), and we are therefore
justified in neglecting them, allowing the other terms to share the
burden of balancing $WS$. We note that this conclusion is justifying
a-posteriori the statement after Eq. (\ref{xy-eq}) that the Cauchy-Schwartz inequality
is saturated for the case considered.

An additional way to see that the cancellation must take place is to examine again the 
momentum balance equation (\ref{monabl}). As discussed below, the MDR is obtained
when the Reynolds stress term $W$  is negligible compared to the polymer contribution
$c_1 \nu_p R_{yy} S$. But when this happens it means that 
\begin{equation}
WS\ll  c_1 \nu_p R_{yy} S^2 \approx c_1\nu_p K/y^2 \ .
\end{equation}
Evidently this means that also in the energy balance equations  $WS$ would be overwhelmed
by terms of the order of $K/y^2$ which therefore must cancel against each other.

Using our order of magnitude estimates for the remaining terms we can therefore conclude
that
\BE \label{inter1}
\ve\sp p (y)\approx  c_2\nu\sb p R_{yy}(y) K(y)/y^2\,,
\EE
where $c_2$ is another parameter of the order of unity. Returning
to the balance equation for the energy, we recall that we cannot
calculate $\ve\sp{dis} (y)$ exactly, but we can estimate it rather
well at a point $y$ away from the wall. When viscous effects are
dominant, this term is estimated as $\nu (a/y)^2 K(y)$ (the
velocity is then rather smooth, the gradient exists and can be
estimated by the typical velocity at $y$ over the distance from
the wall). Here $a$ is a constant of the order of unity. When the
Reynolds number is large, the viscous dissipation is the same as
the turbulent energy flux down the scales, which can be estimated
as $K(y)/\tau(y)$ where $\tau(y)$ is the typical eddy turn over
time at $y$. The latter is estimated as $y/b\sqrt{K(y)}$ where $b$
is another constant of the order of unity. Together with Eq.
(\ref{inter1}) we can thus write the energy balance equation at
point $y$ as
\begin{equation} \label{EB}
a \nu_0 \frac{K(y)}{y^2} +b \frac{K^{3/2}(y)}{y} +c_2 \nu\sb p R_{yy}(y) \frac{K(y)} {y^2}=
W(y)S(y) \ .
\end{equation}
We recognize the important result that the effective viscosity
induced by the rodlike polymers in both the momentum and the energy
balance equation is proportional to $R_{yy}$. \emph{These balance
equations are identical in form to those found for flexible
polymers}~\cite{04LPPT}; this is an important step in understanding the
``Additive Equivalence" discovered by Virk.

To complete the derivation one adds to the balance equation the
relation between $K(y)$ and $W(y)$ which in the elastic layer are
expected to be proportional to each other,
\begin{equation}
K = c\Sb{V}^2 W \ . \label{KW}
\end{equation}
 It should be stressed that rigorously one can establish this
 relation only as an inequality with $c\Sb V\le 1$, and its use as an
 equality (which is common to the derivation of the Newtonian
 log-law as well as to the derivation of the MDR in the flexible
 polymer case) rests on experimental and simulational
 confirmation.  Near the MDR  the terms representing the effect of
 the polymers in Eqs. (\ref{monabl}) and (\ref{EB}) are dominant,
 and one estimates from the momentum equation $R_{yy} (y)\propto
 1/S(y)$. Using this in Eq. (\ref{EB}) together with Eq.
 (\ref{KW}) one ends up with the prediction that $S(y)\propto
 1/y$, leading to a logarithmic law for the mean velocity.
 Repeating the derivation of \cite{04LPPT} in wall units one ends
 up with the MDR Eq. (\ref{final}), with the identification
\begin{equation}
 \kappa\Sb V={c\Sb{V}}/{c\Sb N}{y^+_ v} \ .
\end{equation}
In this equation $c\Sb N$ and $y^+_v$ are constants that appear in
the Newtonian theory, and cannot change from flexible to rodlike
polymers. The existence of drag reduction is guaranteed, since
$\kappa\Sb V$ was shown to be larger than its Newtonian counterpart
$\kappa\Sb K $ \cite{04LPPT}. The actual value of the slope at the MDR
logarithmic law depends nonetheless on the numerical value of
$c\Sb V$. Thus the prediction of the theory is that {\em  if $c\Sb V$ is
about the same in rodlike and flexible polymers,  than the slope of
the MDR should be about the same}.
\section{\label{conclusions} Conclusions}

We have presented a scenario to rationalize the ``additive
equivalence" discovered by Virk and coworkers. The main conclusion
of this paper is that although, on the face of it, the dynamics of
flexible and rodlike polymers appear different, with flexible
polymers being able to ``stretch" and ``store" energy (something
that many researchers thought is central to drag reduction), the
effective Reynolds balance equations for momentum and energy are
isomorphic. Accordingly, the MDR is expected to be the same as
long as $c\Sb V$ of Eq. (\ref{KW}) is the same.

One should stress however that $c\Sb V$ is not expected to be a
universal number. It appears to us unlikely that $c\Sb V$ remains
{\em exactly} the same in rodlike and flexible polymers. It appears
more likely that these respective numbers are of the same order,
giving the impression that the MDR asymptotes are the same. In
addition, one expects that the cross-over from the MDR to the
Newtonian plug, which is non-universal even in flexible polymers
\cite{04BDLPT,04BLPT}, may show significant differences between
rodlike and flexible polymers. Indeed, in friction coordinates drag
reduction by rodlike polymers appears as an upper bound on the drag
reduction by flexible polymers \cite{97VSW}. According to the
theory of \cite{04BLPT} flexible polymers  reach their maximal
drag reduction when fully stretched, being then as effective as
rodlike polymers. This is one way of  rationalizing the findings of
\cite{97VSW}.

To make the difference between the flexible and rodlike polymers
sharper, we note the different $y$ dependence of $K(y)$ and $W(y)$
in the two cases. In the flexible case one had a threshold
condition for the onset of drag reduction in terms of the Deborah
number, stating that the typical time scale for turbulent
fluctuations, $y/\sqrt(K(y)$ is of the order of the polymer
relaxation time $\tau\sb p$. This immediately leads to the estimate
$K(y)\sim y^2$, and the same for the Reynolds stress. In the
present case we have estimated $SR_{xy}\sim \sqrt{K}/y$, and with
$R_{xy}^2\sim R_{yy} \sim y$ we get $K(y)\sim y$, and due to Eq.
(\ref{KW}) we can write
\begin{eqnarray}\nonumber
K(y)&\sim& W(y)\sim y \  , \quad    \text{ for rigid rodlike polymers;}\\
K(y)&\sim&  W(y)\sim y^2 \,  , \quad \text{for flexible polymers.}
\end{eqnarray}
We note that this last statement is a sharp prediction for an
important difference between the two drag reducing universality
classes, a difference that is not at all in contradiction with the
``additive equivalence". A-posteriori we can also see why the terms
of the order of $K(y)/y^2$ in Eq. (\ref{exp-1}) must have
cancelled, being divergent as $1/y$ against a $y$ independent
energy input $W(y)S(y)$. We hope that this prediction would be put
to experimental or simulational test.

An additional important difference between rigid and flexible polymers is that in the latter
case an important condition for attaining the MDR was $R_{xx}\gg R_{yy}$. In the present
case we need to guarantee that $c_1 \nu_p R_{yy} \gg \nu_0$ in order to enable the polymer
terms to overwhelm the Newtonian terms in the balance equations. This condition means 
however that the concentration of the rigid polymer should be large enough before the MDR is
obtained. In the flexible polymer case one could reach the MDR conditions even for small
concentrations as long as the Re is large enough and the Deborah number is large, leading to
$R_{xx}\gg R_{yy}$ \cite{04BDLPT}. This difference leads to the observed experimental behavior, where for flexible polymers the MDR is reached even for small concentrations and then a cross over
back to the Newtonian plug is found, whereas in rigid polymers that MDR is obtained
gradually as the concentration increases, and see the figures in \cite{97VSW} for comparison.

Finally, it is interesting to note that our order of magnitude estimates 
of $R_{ij}$ could be read directly from Eqs. (\ref{orders}) by replacing the 
thermal mean values $\overline{\C R}_{ij}$ with turbulent means 
$R_{ij}\equiv \< \C R_{ij}\> $ and simply identifying the Brownian 
frequency $\gamma\Sb B$ in the definition~\Ref{De} of the Brownian  
Deborah number  $\DE\sb r$ with the characteristic turbulent frequency  
$\gamma\sb {turb}\equiv \sqrt{K} R_{yy}/y$. Put for example in the 
equation for $R_{yy}$ we get
\begin{equation}
R_{yy}\approx 2^{13} K^{1/3}R_{yy}^{2/3}/ [S(y) y]^{2/3} \ .
\end{equation}
Simplifying, this equation reads $S^2 R_{yy} \sim K(y)/y^2$ which
is nothing but the square of Eq.(\ref{est1}). All the other orders
of magnitude derived in Sect. \ref{ss:rules} follow as easily with
this identification. We believe that this is another way to argue
that our estimates are physically sensible and that we have
captured the essence of drag reduction by rodlike polymers and the
nature of the observation of the ``additive equivalence".

\acknowledgments We acknowledge  useful discussions with Anna Pomyalov 
and  Vasil Tiberkevich. This work has been supported in part by the
US-Israel Binational Science Foundation, by the European Commission
under a TMR grant, and by the Minerva Foundation, Munich, Germany.

\end{document}